\documentclass[aps,prd,onecolumn,12pt,preprintnumbers]{revtex4-1}

\usepackage{hyperref}

\usepackage{float}

\usepackage{tikz}

\usepackage{graphicx}
\usepackage{epstopdf}

\begin{document}

\preprint{HDP: 15 -- 04}

\title{Sound Hole Sound}

\author{David Politzer}

%\email[]{politzer@theory.caltech.edu}
\email[]{politzer@caltech.edu}

\homepage[]{http://www.its.caltech.edu/~politzer}

%\thanks{452-48 Caltech, Pasadena CA 91125}
\altaffiliation{\footnotesize 452-48 Caltech, Pasadena CA 91125}
%\altaffiliation{\newline \em \em \em 452-48 Caltech, Pasadena CA 91125}
\affiliation{California Institute of Technology}

%\date{\today}
\date{May 10, 2015}

\begin{figure}[h!]
\includegraphics[width=5.6in]{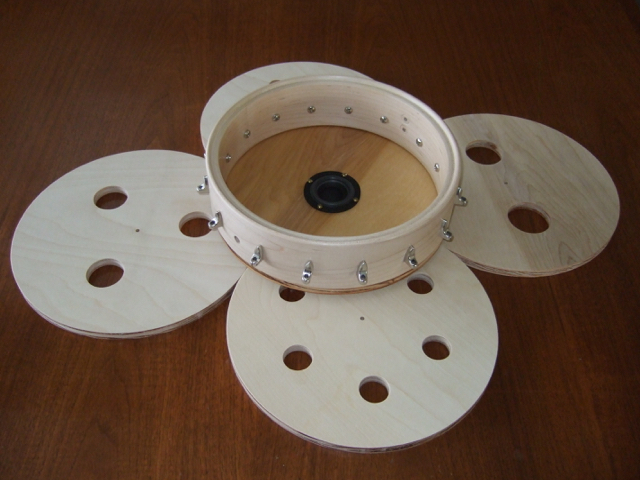}
\end{figure}

\begin{abstract}

The volume of air that goes in and out of a musical instrument's sound hole is related to the sound hole's contribution to the volume of the sound.  Helmholtz's result for the simplest case of steady flow through an elliptical hole is reviewed.  Measurements on multiple holes in sound box geometries and scales relevant to real musical instruments demonstrate the importance of a variety of effects.  Electric capacitance of single flat plates is a mathematically identical problem, offering an alternate way to understand the most important of those effects.  The measurements also confirm and illuminate aspects of Helmholtz's ``bottle" resonator model as applied to musical instrument sound boxes and sound holes.
\end{abstract}

\maketitle{ {\centerline{{\large \bf Sound Hole Sound}}

\section{Introduction}

One mechanism of sound production in stringed instruments is the in and out flow of air through a sound hole, forced by sound vibrations within the sound box.  Helmholtz\cite{helmholtz}  and Rayleigh\cite{rayleigh} studied how the magnitude of that in and out flow depends on sound hole size and shape for a given pressure within the sound box.

Helmholtz solved the simplest possible version of sound hole flow.
For holes of similar shape (i.e., mathematically similar: differing only in overall scale) and under generally plausible simplifying assumptions, the flow through the hole is proportional to its linear dimensions (and the pressure difference between the two sides) --- and {\it not} to the area of the hole.\,(!)   For convex shapes, the flow is primarily near the edge, whose length scales linearly with overall size.  Rayleigh pointed out a consequence.  If we compare a single hole to $N$ similarly shaped smaller holes, such that the total area is the same, the total flow through the $N$ smaller holes will increase like $\sqrt N$, as long as the small holes are sufficiently distant from each other.

This suggested a banjo-based experiment.   Some relevant parts are shown in the picture on the title page.  The physics and acoustics are of general interest and applicability --- but not all that relevant to banjos in particular, with the possible exception of cigar box banjos.  The apparatus included some banjo parts but in ways quite removed from playable instruments.

If we're interested in the sound box of a musical instrument with possibly more than one sound hole, the side walls of the box and the proximity of the several holes to each other both reduce the flow relative to the $\sqrt N$ scaling.  This is apparent in the results of the sound hole experiment and will be explained, at least qualitatively.  The measurements also offer a way to quantify the effective length of the ``neck" in the Helmholtz bottle resonator model of the lowest resonance of the cavity.  This length is unambiguous for a real bottle with a cylindrical neck that is long compared to its diameter.  In the case at hand, there is a plug of air that goes in and out of the sound hole(s) as the sound box air contracts and expands.  The measurements allow an estimate of the plug mass.

The flow through different shape holes of the same area typically grows with their perimeter (as long as the shape is convex and the hole is not so thin that friction becomes a significant factor).  The sound hole on a resonator banjo is the space between the bottom edge of the rim and the back of the resonator.  So its length is the circumference of a roughly $10''$ diameter circle, and its width is a small fraction of an inch.  That's about as much convex circumference as one can practically get for a given hole area on the ``pot" (drum body) of a banjo.  (The hole area is tuned to support the lowest frequencies.\cite{rae})  One might ask: is this geometry of the resonator the key to its loud sound?  In an accompanying paper, I note that the practical banjo alternatives, i.e., open back and flat disk back, have similar geometry sound holes and are similarly loud.\cite{resonator})

\section{Helmholtz's Sound Hole Result}

Rayleigh\cite{rayleigh} calculated the flow of a fluid through an elliptical hole in an infinite plane baffle that separates two half-spaces at slightly different pressures.  He noted that Helmholtz published the result seventeen years earlier.\cite{helmholtz}  No other shapes can be calculated analytically.  Crucial approximations render the calculation doable.  In particular, viscosity is ignored, and the flow is assumed to be irrotational.  Furthermore, the fluid is assumed to be incompressible.

These are audacious claims.  Viscosity, no matter how small, is always important as fluid moves near a solid surface.  For almost all materials, the flow rate goes to zero as you approach the solid.  Dramatic, everyday examples were given by Feynman\cite{feynman}, involving dust and wind.  Fans get dusty, and you can't blow the dust off your car by driving fast.  That is because there is no air motion relative to the solid surface at the surface.  In the gentlest of cases, there is a transition region where viscosity is essential in mediating the flow between the solid surface (where it is at rest) and the fluid interior, where viscosity might no longer be an important factor.  This is known as the laminar boundary layer.  In the sound hole case, Helmholtz and Rayleigh assume that the boundary layer is thin and can be ignored as a first approximation.  The validity of the approximation depends on the how thin the boundary layer is relative to the size of the hole.

Air is certainly compressible.  Furthermore, we're interested in sound, which is waves of compression and rarefaction.  However, Helmholtz and Rayleigh knew to ask a simpler question: what is the steady flow through a hole in a barrier that separates two infinite regions with slightly different pressures.  Even with highly compressible fluids, it often turns out in the end that, for gentle flows, the density change is negligible.  It is then inordinately convenient in doing calculations to imagine that the fluid simply could not be compressed in the first place.  An astounding example of assuming incompressible flow is interstellar and intergalactic gas or dust.  What could be more compressible than a gas with one particle per cubic meter (or less)?  Nevertheless, in some situations the solution found by implementing incompressibility from the start can be seen to be consistent with that assumption in the end, were the assumption relaxed.

A consequence of these assumptions is that the steady fluid velocity, which has a magnitude and direction at each point in space, can be represented by just a number at every point in space, where the relation between the two is given by simple calculus.  (In math terms, the vector velocity field, {\bf v}({\bf r}), is the gradient of a scalar potential $V(${\bf r}$)$, i.e., {\bf v}({\bf r}) = ${\bf \nabla} V(${\bf r}$)$.)  It is far simpler to solve first for that number as a function of position (i.e., the scalar field).  In fact, all that is needed in addition to the assumptions stated thus far,\cite{laplace} is how $V(${\bf r}$)$ behaves on the boundary, i.e., across the sound hole and along the barrier.  Similarly, in electrostatics, the vector electric field {\bf E}({\bf r}) is the gradient of a scalar potential.  A precise electrostatic analog problem to sound hole flow is the capacitance of a single plate capacitor, i.e., in the shape of the sound hole.  The equations for the scalar potential, including the boundary conditions, are identical.  Hence, they have the same solutions.  A version of the electrostatics calculation is presented in many advanced texts on electromagnetism\cite{jackson} (though Rayleigh's is simpler than most).  In the fluid flow language, the boundary conditions are that the flow along the baffle must be parallel to the surface (and not into it) and (by symmetry) that the flow is straight through the hole and not at some other angle or angles.

Pioneers in the theory of electromagnetism used analogies with fluid flow in their thinking and presentations.  By the $20^{\text{th}}$ Century, it became the other way around.  So I will discuss the problem mostly in terms of electric charge and capacitance.  Capacitance is the ratio of the charge on a capacitor to its voltage difference (electric potential).  Sound hole conductance is the ratio of the total flow rate through the hole to the pressure difference between the two regions on the two sides of the baffle, far from the hole.
   
In Gaussian cgs units, capacitance is a length, measured in centimeters.  The conversion to more common SI MKS units is that 1 cm of capacitance is approximately 1 pF.  In college physics courses, one encounters parallel plate capacitors with plate area $A$ that is much larger than their separation-squared $d^2$; that capacitance 
is ${1 \over {4 \pi}} A / d$.  (There are extra constants in SI units that confuse considerations of dimensional analysis.)  Another commonly discussed problem is the isolated conducting sphere of radius $R$.  Its capacitance, where the potential difference is defined to be relative to infinitely far away, is simply $R$.  More generally, the capacitance of a single conducting object relative to infinity must be some linear measure of its size times a dimensionless function of its shape. So, a family of geometrically similar objects will have capacitances proportional to that linear measure times the same dimensionless function that comes from their common shape.  Coaxial cylinders, e.g. coax cable, have a capacitance per unit length of $(2 \text{log} {b \over a})^{-1}$, where $b$ and $a$ are the outer and inner radii.  (From this, one can conclude that a single infinitely long conductor does not have a finite capacitance per unit length.)

The fluid conductance of an isolated, single hole is what Helmholtz and Rayleigh calculated for the shape of a generic ellipse.   For the special case of a circle of radius $R$, the fluid conductance is $2R$.  This is the same problem as the capacitance of an isolated circular disk (up to a factor of $\pi$). The case of the general ellipse is ``solved" in that it is reduced to ``quadrature."  It involves  special functions that have been well studied and tabulated over the years by applied mathematicians but cannot be expressed exactly in terms of elementary functions, i.e., the ones we're supposed to learn about in high school.  Rayleigh\cite{rayleigh} considered it worth tabulating the numerical results for a range of eccentricities of ellipses all of the same area.  He emphasized that the fluid flow (or capacitance) changes surprisingly slowly as the shape becomes more eccentric.

So, single plate capacitance and sound hole conductance are proportional to their linear size and not their area.  The ``cause" is easy to visualize.  In the case of electric charges on a conducting plate, the charges repel each other.  Their equilibrium, static distribution is a compromise between them getting as far away from each other as possible, which favors piling up on the (convex) edge, and not all of them going right to the edge --- because they repel each other.  But not much charge is left in the middle.  To think about the sound flow, visualize the velocity streamlines.  There is certainly some flow straight through the middle of the hole.  But material originally far from the hole in the lateral  direction (parallel to the baffle) will converge towards the hole as it approaches from the high pressure side.  There will be a greater density of streamlines near the edge of the hole.  So the flow is fastest there.

\section{multiple holes at finite separation and proximity to walls}

If there are multiple sound holes on a single instrument, they cannot be infinitely far apart.  They will necessarily reduce each other's flow.  Also, because of the finite size of the sound box, the proximity of the side walls will similarly also reduce the flow through the holes.   In both cases, there's just less stuff per hole.  These effects can be quantified in relevant, limiting cases --- but it is still true that any sound hole problem besides the isolated, single, ellipse cannot be solved analytically.  

The electric language leads directly to a quantitative treatment of simple, idealized situations.

Imagine $N$ identical conducting disks, each carrying a charge $Q/N$.  Initially the disks are so far from each other that we can ignore their interaction.  Then the total capacitance is just $N$ times the capacitance of a single such disk.  If we now bring them closer together, we do work against their repulsion.  So we are putting energy into the system.  The total electric energy $U$ of a capacitor with total charge $Q$ is $U = {1 \over 2} Q^2 / C$, where $C$ is its capacitance.  If we've increased $U$ at fixed $Q$, $C$ must have decreased appropriately.

A simple calculation illustrates this effect.  Consider $N$ circular disks with {\it total} area $\pi R^2$ and {\it total} charge $Q$.  Compare $N=2$ to $N=4$.  Begin with the disks infinitely separated.  The capacitances in the two cases are $C^\infty_{N=2}=2{\sqrt 2} R/\pi$ and $C^\infty_{N=4}=4R/\pi$.  Now bring the disks in from infinity so that they lie on a circle of diameter $d$.  The two $N=2$ disks are separated by $d$.  The $N=4$ disks are at the corners of a square with diagonal $d$.  As long as $d \gg R$ and the rearrangement of charge on each disk can be ignored, one can easily calculate the work done to bring the disks closer.

For $N=1$, a single disk of radius $R_1$ has a capacitance $C_1 = 2 R_1/\pi$.  For $N=2$, with $R_2$ such that the total area is the same as the single $R_1$ disk, when the two are brought to a separation $d$, with $d \gg R_2$, the capacitance is $C_2 \approx ({\pi \over {2\sqrt{2} R_1} }+ {1 \over {2d}})^{-1}$.  The ${1 \over {2d}}$ term reduces the capacitance by an amount that reflects the work done in bringing the two disks from infinity to a separation of $d$.  Assembling the $N=4$ configuration as described, $C_4 \approx ({\pi \over {4 R_1} }+ {{\sqrt{2} + 1/2} \over {2d}})^{-1}$.

Some level of quantification of the effect of proximity to a side wall is given by the following.  In the fluid language, the two sides of the baffle need not be identical.  They only have to match across the hole.  In the electric case, the field lines reverse direction across the dividing plane, and the two sides are otherwise symmetrical.

In the fluid case, if there is a ``side" wall near the hole (a plane that is perpendicular to the original, infinite baffle), flow will be along the wall, and no material will cross it.  Hence, the flow with the side wall in place is the same as the flow with no side wall but with a second identical sound hole located equidistant from where the side wall had been.

In the electric language, this is akin to the method of images for the static effect of a conducting plane, except that here the image has the {\it same} charge as the original disk.  And I previously described how the presence of nearby additional holes or disks reduces the flow or capacitance.  (You don't count the flow through the image, but only its reduction of flow though the original hole.)

\section{measurements of sound hole sound conductance}

\begin{figure}[h!]
\includegraphics[width=5.0in]{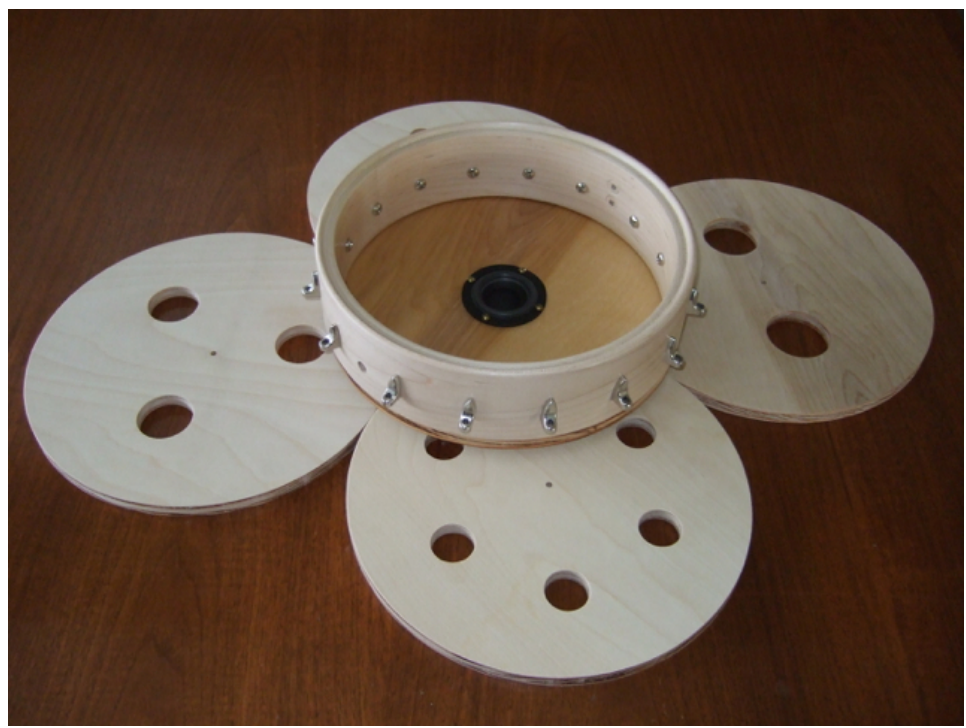}
\caption{Speaker, cavity, and baffles with holes used to measure conductance}
\end{figure}

I used an $11''$ banjo rim for the side walls of a cavity, shown in FIG.~1.  In place of the banjo drum head, there was a $1/2''$ thick plywood disk with a full-range $2''$ speaker mounted at its center.  For the other side of the cavity, I mounted one of several $1/2''$ plywood disks.  These disks had 1 to 8 symmetrically placed holes, sized so that the total hole area was the same in all cases.

(These measurements have no real relevance to actual banjos, where the drum head is an essential component and the sound holes cannot be on the front or back.  In practice, standard banjo sound holes already have an extreme geometry, i.e., they are ($10''$ diameter) rings.\cite{resonator})

I drove the speaker with an impedance-matched signal generator though a range of frequencies that spanned the range of Helmholtz resonances.  The frequencies of the Helmholtz resonances could be identified roughly in the traditional way by blowing compressed air across the holes, but that procedure is very noisy.  (Note that, in the absence of a flexible drum head or sound board, there is no nearby, strongly coupled resonance corresponding to the lowest motion of the sound board.)

The focus on the Helmholtz resonance is because it produces flow through the holes that is simultaneously either all inward or all outward.  The Helmholtz resonance pressure is the same at the locations of each of the sound holes.  Hence, the sound as measured by a microphone nearby (in particular, at $11''$ from the plane of the holes), placed symmetrically, i.e., on axis, is a measure of the total flow in response to the speaker driving.  The sound from each hole travels the same distance to the microphone.  So the contributions from the several holes arrive in phase and add directly.  The resonant frequencies for different $N$ are relatively close to each other.  The voltage amplitude is constant over the sweep.  So it is assumed that the speaker driving amplitude is the same for all cases.

(The observed resonant frequencies are somewhat different for different $N$.  This is of interest, and is addressed subsequently.)

FIG.~2 displays the results for $N$ = 1, 2, 3, 6, and 8.  The loudness appears to increase steadily with increasing $N$.  From $N=1$ to $N=8$ the increase is about 3 dB, which is a factor of $\sqrt{2}$ ($\approx 1.4$) in pressure at the microphone.  Were the 8 holes infinitely separated on an infinite baffle, we would expect the increase in sound pressure (if we could somehow add the 8 hole pressures) to have been $\sqrt{8} \approx 2.8$ --- because the external sound pressure scales linearly with the amount of air going in and out.  To fit on a musical instrument, the sound holes cannot be infinitely separated, and they are not infinitely far from the side wall.  As discussed above, both of these facts reduce the flow relative to infinitely separated holes and more so with increasing $N$.  A proper quantitative calculation of a realistic geometry, e.g., in terms of charged plates, would require allowing the charge distribution to relax to equilibrium on each plate.

\begin{figure}[h!]
\includegraphics[width=6in]{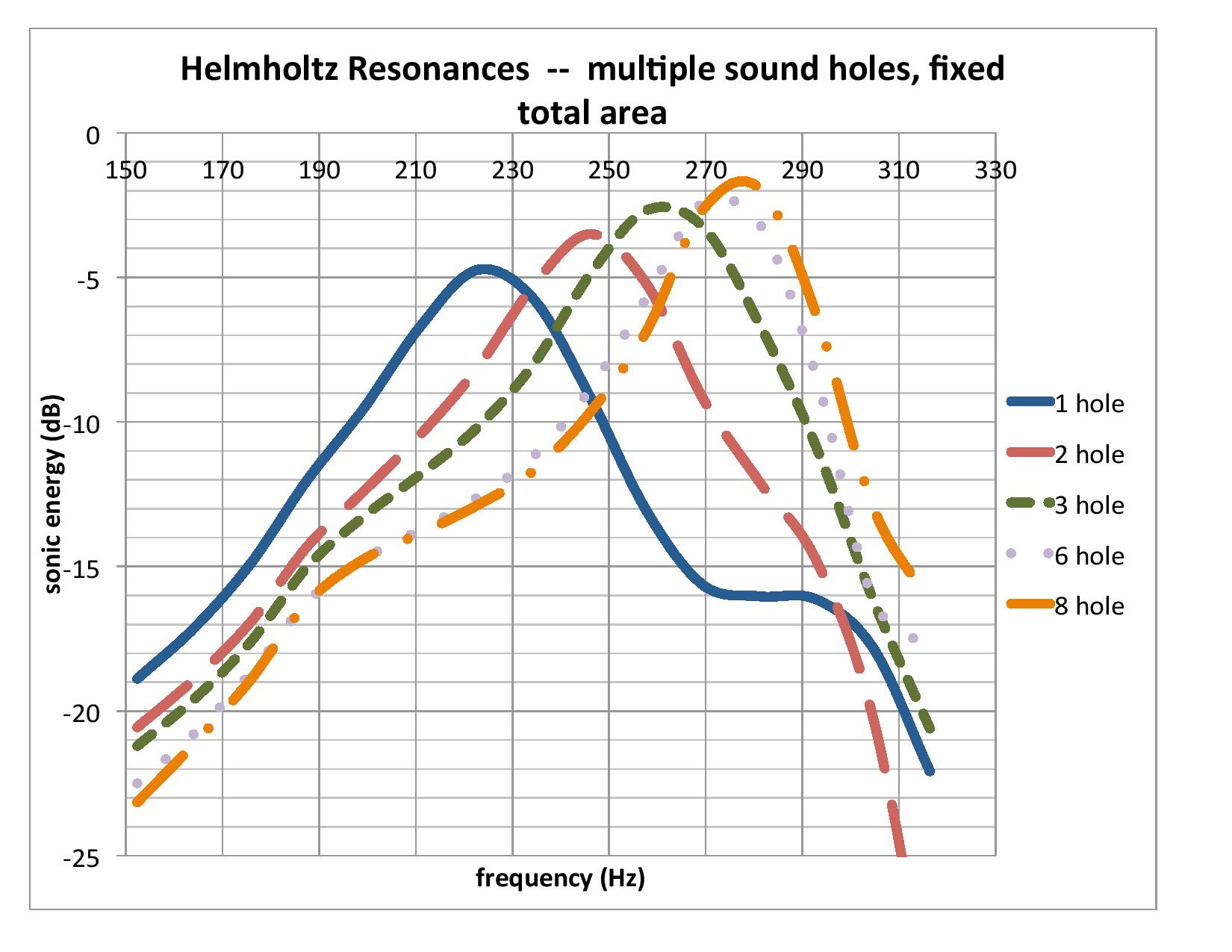}
\caption{Helmholtz resonances for 1, 2, 3, 6, 8 circular sound holes with same total area}
\end{figure}

\section{Effective Neck Length in the Helmholtz Resonance Frequency}

\begin{figure}[h!]
\includegraphics[width=2.5in]{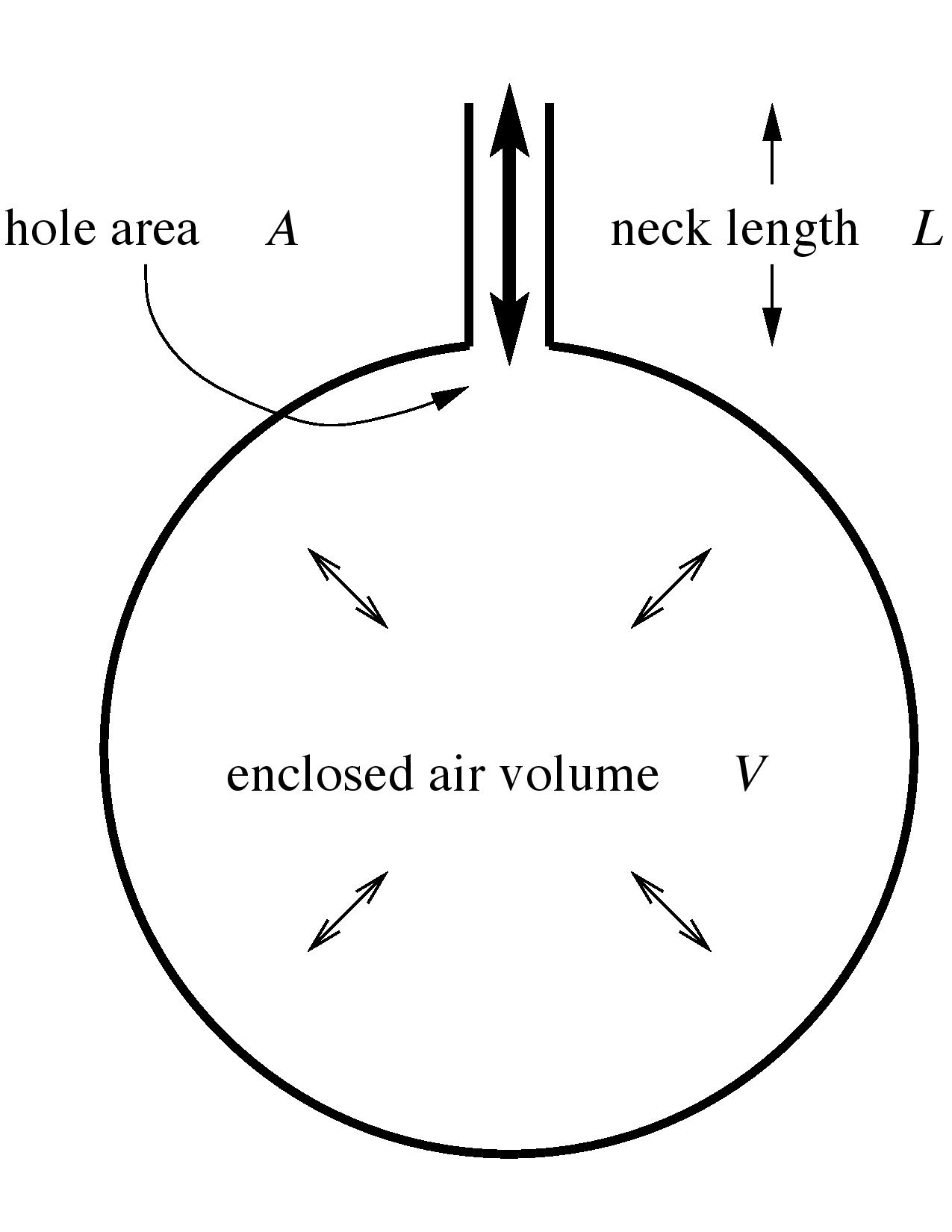}
\caption{Ideal Helmholtz resonator}
\end{figure}

Helmholtz's estimate for the bottle resonant frequency is $f_{\text{Helmholtz}} = {v_s \over {2 \pi}} \sqrt{ A / (V L)}$, where $v_s$ is the speed of sound, $L$ is the length of the cylindrical bottle neck, $V$ is the bottle volume, and $A$ is the cross sectional area of the neck, all as appear in FIG.~3.

For the banjo pot used in these measurements, $V$ and $A$ can be measured directly.  $V \cong \pi (9.82''/2)^2 \times 2.76''$; and $A \cong \pi (2.70''/2)^2$.  Actually, there is a couple of percent difference in $A$ for each $N$ because of the realities of using readily available hole saws and the cuts they make; the final calculations and comparisons include the measured values for each $N$.  In particular, the measured hole diameters $D_N$ were $2.70''$, $1.96''$, $1.59''$, $1.13''$, and $0.95''$ for $N$ = 1, 2, 3, 6, and 8, respectively.  Let $A_N \equiv \pi (D_N / 2)^2$.  Then the total hole areas $N A_N$ were 5.73, 6.03, 5.96, 6.02, and 5.67 in$^2$ for $N$ = 1, 2, 3, 6, and 8, whose extremes deviate from the average by about $\pm 3\%$.

The thickness of the plywood is actually $0.475''$, but $A L$ in the ideal formula for $f_{\text{Helmholtz}}$ is the volume of the air being moved in and out by the oscillating expansion and contraction of the air in the cavity.  In the derivation, the $L$ in the denominator started out as $A L$ (the volume of the neck), and the $A$ in the numerator was $A^2$.  For this model to make sense, it is essential that $A L \ll V$.  To apply this formula in the present situation, $L$ must be an effective length that includes in the volume $A L$ the air inside and outside that actually moves substantially as the air enclosed in $V$ expands and contracts.  $A$ has a dual significance in the ideal Helmholtz geometry.  $AL$ is the neck volume, and $A$ is the area over which the pressure variations of the big volume act on the neck volume.  By putting the neck mass corrections into an effective $L$, we retain the role of $A$ as the interface area.

The observed frequencies $f_N$ for $N$ = 1, 2, 3, 6, and 8 were 225, 246, 261, 273, and 278 Hz, respectively, as displayed in FIG.~2.

There are many ways one might try to compare these measurements
to the Helmholtz resonator equation.  It is commonly stated is that the effective length $L_{\text{eff}}$ is the sum of the actual hardware length and a number times the hole diameter.\cite{L-eff}  For pipes that are long compared to their diameters and open at each end, the effective length for the lowest pipe resonance is often parameterized as $L_{\text{eff}} \approx L_o + \alpha D$, where $L_o$ is the actual pipe length, and $D$ is its diameter.  $\alpha$ is measured to be around 0.35.  For the case at hand, $D$ is certainly bigger than $L_o$, but one may parameterize the results similarly and ask what effective hole length would be needed to fit the observed frequencies to the simple Helmholtz resonator formula.  And finally, one can express the extra length as a number times the hole diameter and do that separately for each value of $N$, the number of holes.  (The actual holes in the measurements varied in diameter by a factor of 2.8.)  Putting all this into equations:

\medskip

\centerline{$f_N = {v_s \over {2 \pi}} \sqrt{ (N A_N)^2 / (V N A_N L_N)} = {v_s \over {2 \pi}} \sqrt{ ( N A_N / (V  L_N)}$}

\noindent and 

\centerline{$L_N = L_o + \beta_N D_N$ .}

\noindent $f_N$ are the frequencies; $A_N$ are the individual hole areas (which are all very nearly equal to $A_1 / N$), and $D_N$ are the individual hole diameters (all nearly equal to $D_1 / \sqrt{N})$; $L_o$ is the wood thickness; and $\beta_N$ is the neck length correction factor relative to the individual hole diameter, to be deduced for each $N$ from the measured $f_N$'s.

The values for $\beta_N$ produced by this exercise are 0.75, 0.88, 0.92, 1.17, and 1.22 for $N$ = 1, 2, 3, 6, and 8, respectively.  It is certainly not the case that $\beta_N$ is a number independent of $N$.  Rather, there seems to be a systematic trend to increase with  $N$.  However, the numbers are all within a factor of 2.  They are not only plausible, they are certainly in the vicinity of the value 0.82 estimated by Rayleigh,\cite{rayleigh} who naturally considered a single, isolated hole. 

The increase in the fit values of $\beta_N$ with increasing $N$ is plausibly attributable to the increasing proximity of the holes.  Nearby holes may be cooperating with each other to get the air moving.  So even more air is moving (and contributing to the mass term in the Helmholtz oscillator formula) than the sum of the holes singly.  It is also possible that the $N = 1$  value falls below Rayleigh's number because of the presence of nearby side walls, which reduce the flow in their vicinity. 

A parameterization that gives a nearly perfect fit to the measurements, albeit without any theoretical justification, is given by

\centerline{$L_N = L_o + b_N \sqrt{D_N}$}
\noindent to define $b_N$.  From the measured frequencies, $b_N$ = 1.24, 1.24, 1.16, 1.24, and 1.19 in$^{1 \over 2}$ for $N$ = 1, 2, 3, 6, and 8.  Such a form would only be possible if there were a second length parameter, beside the hole diameter, that were relevant.  In this case, the hole separation and/or wall distance might play that role. 

A numerical simulation of the flow might give a hint as to the shape of the oscillating disturbance, how it scales with hole size, and how neighboring holes interact.


\begin{thebibliography}{99}
\bibitem{L-eff}Rayleigh[2] makes estimates and bounds of this form.  One might argue that for a single, isolated hole in an infinite baffle under the simplifying assumptions used to find the fluid flow conductance, there is only one relevant length parameter, i.e., the diameter $D$ of the hole.  In the oscillating case of interest there is a second length: $\lambda$, the wavelength of sound at the driving frequency.  But the Helmholtz resonator equation only makes sense for the linear dimensions of $V$ to be much smaller than $\lambda$, and the dimensions of the hole, e.g., $D$, must be much smaller still.  In the present situation $\lambda$ is so much bigger than $D$ that it can be safely ignored.  Were we interested in the disturbance near a hole for $\lambda \ll D$, the situation would be quite different.  The hole would be radiating a beam.
\end{thebibliography}
\end{document}